\documentclass[preprint2]{aastex}
\usepackage{graphicx}
\usepackage{natbib}

\shorttitle{Star Formation Activity in the Pipe Nebula}
\shortauthors{Forbrich et al.}

\begin{document}

\title{A \textit{\textbf{Spitzer}} Census of Star Formation Activity in the Pipe Nebula}

\author{Jan Forbrich, Charles J. Lada, August A. Muench}
\affil{Harvard-Smithsonian Center for Astrophysics, 60 Garden Street, Cambridge, MA 02138}
\email{jforbrich, clada, amuench@cfa.harvard.edu}
\and
\author{Jo\~{a}o Alves}
\affil{Calar Alto Observatory, Centro Astron\'omico Hispano Alem\'an, C/Jes\'us Durb\'an Rem\'on, 2-2, 04004 Almeria, Spain}
\email{jalves@caha.es}
\and
\author{Marco Lombardi}
\affil{European Southern Observatory, Karl-Schwarzschild-Str. 2, 85748 Garching, Germany}
\email{mlombard@eso.org}

\begin{abstract}
The Pipe Nebula, a large nearby molecular cloud lacks obvious signposts of star formation in all but one of more than 130 dust extinction cores that have been identified within it. In order to quantitatively determine the current level of star formation activity in the Pipe Nebula, we analyzed 13 square degrees of sensitive mid-infrared maps of the entire cloud, obtained with the Multiband Imaging Photometer for Spitzer (MIPS) at wavelengths of 24~$\mu$m and 70~$\mu$m to search for candidate Young Stellar Objects (YSOs) in the high-extinction regions. We argue that our search is complete for class~I and typical class~II YSOs with luminosities of $L_{\rm bol}\sim0.2$\,$L_\odot$ and greater. We find only 18 candidate YSOs in the high-extinction regions of the entire Pipe cloud. Twelve of these sources are previously known members of a small cluster associated with Barnard\,59, the largest and most massive dense core in the cloud. With only six candidate class~I and class~II YSOs detected towards extinction cores outside of this cluster, our findings emphatically confirm the notion of an extremely low level of star formation activity in the Pipe Nebula. The resulting star formation efficiency for the entire cloud mass is only $\sim0.06$\%.
\end{abstract}


\section{Introduction}

The most critical but least understood aspects of the star formation process concern its initial conditions and earliest stages. In particular, little is known about the initial phases of protostellar development within dense molecular cloud cores and even less is known about the origin of the dense cores themselves.  This is largely due to an apparent absence of nearby ($d<2.5$\,kpc) massive ($>10^4$\,$M_\odot$) molecular clouds in a sufficiently early evolutionary state so that little or no star formation activity is present within them. This is then taken as evidence that the onset of star formation in molecular clouds is extremely rapid and that clouds in the earliest evolutionary states must be exceedingly rare. Clearly the identification and detailed study of such clouds provides invaluable information about the earliest phases of star formation.

With a mass of 10$^4$\,$M_\odot$ and a distance of 130~pc \citep{lom06}, the Pipe Nebula is one of the massive clouds nearest to the Sun and yet has received little attention and has only been studied in detail within the last few years. With the exception of Barnard 59, the region is lacking the most obvious signposts of star formation, such as bright H\,{\sc \small II} regions, reflection nebulae, luminous infrared sources and large populations of emission-line stars and may be a prime example of a molecular cloud in an early stage of evolutionary development. As such, it constitutes a critical laboratory for star formation studies (for a review, see \citealp{all08}). The first wide-area study of the region consisted of CO mapping carried out by \citet{oni99} who identified 14 C$^{18}$O cores in a large low-resolution and undersampled map. Subsequently, \citet{lom06} presented a high-resolution completely sampled dust extinction map (based on 2MASS near-infrared data) to study general properties of the cloud. This extinction map was also used to study its dense core population: \citet{alv07} identified a population of roughly 150 extinction cores and found that the shape of the core mass function (CMF) is nearly identical to the stellar initial mass function (IMF), but shifted to higher masses by a factor of three, suggesting that the CMF is a direct evolutionary precursor to the stellar IMF. Most recently, \citet{rat09} have further refined the core identification by taking into account additional measurements of C$^{18}$O strength and velocity towards single cores as indicators of dense material and cloud membership. Their list contains 134 dense cores in the Pipe Nebula. To more reliably detect the extinction peaks, \citet{rat09} have used a background-subtracted version of the near-infrared extinction map, produced by filtering out structures surpassing a given size (see also \citealp{alv07}). Note that in the following, we we will generally refer to this background-subtracted map as the extinction map, except as otherwise noted.

The only previously known region of active star formation in the Pipe Nebula is the dark cloud Barnard~59 \citep[B\,59,][]{bar27}. Early studies found indications of ongoing star formation in B\,59: H$\alpha$ emission line stars \citep[][and references therein]{her05}, a millimeter continuum source \citep{rei96}, and a CO outflow \citep{oni99}. More recently, \textsl{Spitzer} observations revealed a young cluster inside of B\,59 \citep{bro07}. \citet{ria09} studied the cluster's most deeply embedded object, [BHB2007] 11, in more detail. In order to quantify the star formation activity for the whole cloud complex, we present a mid-infrared survey carried out with the Multiband Imaging Photometer for Spitzer (MIPS) as well as with the Infrared Array Camera (IRAC), both onboard the \textsl{Spitzer} Space Telescope.

\section{Observations and Data Reduction}

We present large-area mapping observations of the Pipe Nebula, carried out with \textsl{Spitzer}-MIPS \citep{rie04}, at wavelengths of 24~$\mu$m and 70~$\mu$m. The observations were obtained between April 2004 and October 2006. The data consist of 44 Astronomical Observation Requests (AORs). They come from three different programs: the Cycle 1 GTO observations of B68 (Program Identification PID 53, 1 AOR), six mipsphot observations from Legacy observations (PID 139, c2d) and 35 mipsscan as well as 3 mipsphot maps from the Cycle 2 GO Pipe Cloud Survey (PID 20119). We use \textsl{Spitzer}-IRAC data that were mostly collected as part of program PID~20119, but also as programs PID~139 and PID~30570.
The observations were designed to cover the most highly extincted regions of the Pipe Nebula. In total, the 24~$\mu$m data cover an area of 13.4 square degrees, with the 70~$\mu$m data covering a very similar area (to within 2\%). 

The data reduction was performed with the MOPEX software package, v18.1.5\footnote{http://ssc.spitzer.caltech.edu/postbcd/mopex.html}. The 24~$\mu$m data were self-calibrated by dividing each Basic Calibrated Data (BCD) image with a flatfield created by stacking
all the BCDs in an AOR. All MIPS BCDs were from release S16. Self calibration in this way removes large scale sky variations and some artificats. The flat field image was created and applied using the flatfield.pl tool provided as part of MOPEX. Briefly, the BCD stack are median filtered in order to detect sources, mask bright sources, and then median stack the original frames into a flat, clipping out the bright sources. No overlap correction was applied to the 24~$\mu$m data since flatfielding appeared to be sufficient. The AORs were individually processed into mosaics to preserve the two-epoch nature of most of the 24~$\mu$m data, allowing the rejection of asteroids in the final source catalog. The mosaicking produced images with a final pixel size of 2.5$''$ and included steps to mask faulty bits and reject outliers. No self-calibration or overlap corrections were applied to the 70~$\mu$m BCDs where we work with the median-filtered (fbcd) frames produced by the Spitzer Science Center pipeline process. The pixel size at 70~$\mu$m is 4$''$. 

Source detection was performed on the individual 24~$\mu$m mosaics, using MOPEX multi-frame PRF-fitting photometry. Given the Pipe Nebula's location close to the ecliptic and thus an increased probability to detect asteroids, we required source detections in both of the two observation epochs, if two observing epochs were available. The concatenated list of source detections contains nearly 80000 sources above a signal-to-noise ratio of four. In contrast, the 70~$\mu$m map mostly has only single-epoch coverage. Source detection was carried out on a single, large map produced from the median-filtered BCD files. In this map, a total of 5308 sources were detected above a signal-to-noise ratio of four. A cross-correlation of the two lists with a search radius of 4$''$ yields a list of 546 sources detected in both MIPS bands.

For the identified candidate YSOs, we obtained \textsl{Spitzer}-IRAC mid-infrared aperture photometry from corresponding post-basic calibrated data (PBCD) sets for use in the fitting of their spectral energy distributions (SEDs), allowing a photometric accuracy of ~$\sim10$\%. Post-BCD mosaics from the Spitzer Science Center (pipeline S14) were used without modification. Aperture radii used were either two or three pixels (the former only for sources 22 and 24 to exclude neighboring sources), with a pixel size of 1.2$''$ and the corresponding aperture corrections. In a few cases, this required the use of a tool\footnote{http://spider.ipac.caltech.edu/staff/jarrett/irac/tools/} to rectify saturated sources.

While we processed and analyzed the crowded young cluster region Barnard~59 in the same way as all other data, we make use of the IRAC and MIPS photometry from \citet{bro07} for two reasons: first, the crowding is especially problematic in the MIPS data that we use for source detection, causing us to miss a few sources. However, no comparable crowding occurs in the entire remaining areas in the Pipe Nebula. Second, the most clearly saturated sources are in B\,59, requiring more effort to extract reliable photometry. As was pointed out by \citet{bro07}, the three brightest 24~$\mu$m sources in B\,59 ([BHB2007] 1, 7, and 11) are saturated while at 70~$\mu$m, only [BHB2007]~11 is saturated. Outside of B\,59, among the sources we are studying, only KK~Oph is saturated at 24~$\mu$m.

In addition to crowding of bright sources, our reliance on MIPS data for the identification of YSOs also suffers from the relatively low angular resolution: the detectability of multiple systems is reduced even when compared to 2MASS. For example, KK~Oph is a known binary with a separation of 1.6$''$ (see below), but we cannot resolve the two components.

\section{Analysis}

\subsection{Selection of candidate YSOs}
\label{sec_selec}

We expect to find any candidate protostars in the relatively high-extinction regions of the cloud. These regions correspond to only a small percentage of the full coverage. In line with earlier analysis \citep[][in particular]{rat09}, we use the following threshold to define the high-extinction regions: in an extinction map in which extended structure has been removed by means of wavelet subtraction \citep{alv07} we use a minimum extinction of $A_V>1.2$~mag. An area of 0.98~square degrees is above this threshold out of which the identified dense cores \citep{rat09} comprise an area of 0.75~square degrees. In the following, we will constrain the star formation activity in these high-extinction regions. It will become clear that we can constrain the sizes of the populations of both class~I protostars and class~II YSOs (as defined by \citealp{lad87}) in the high-extinction regions.

Given the location of the Pipe Nebula, projected onto the vicinity of the Galactic center, the background source density in the region is enormous and selection criteria for candidate YSOs are of particular importance. Compared to source detection at near-infrared wavelengths, the lower source density in the mid-infrared at 24~$\mu$m and 70~$\mu$m allows us to work with a sensitivity-limited (rather than confusion-limited) sample of objects. When looking for candidate protostars at these wavelengths, we first have to establish whether the mid-infrared mosaics are sensitive enough for this purpose. A good first test is to look at the lowest-luminosity protostellar objects known, the Very Low Luminosity Objects (VeLLOs). These objects were only discovered recently, with the unprecedented sensitivity of \textsl{Spitzer}. The four VeLLOs that have been studied in detail \citep[][, L1014-IRS, L1521F-IRS, IRAM 04191, and L328-IRS, respectively]{you04,bou06,dun06,lee09} have been detected with MIPS at both 24~$\mu$m and 70~$\mu$m, and their mean bolometric luminosity is 0.25~$L_\odot$ (the extreme values being 0.15~$L_\odot$, IRAM 04191, and 0.36~$L_\odot$, L1521F-IRS). Note that the bolometric luminosities are significantly higher than the intrinsic luminosities of the corresponding central objects. Detections at both 24~$\mu$m and 70~$\mu$m are also used as criteria by \citet{dun08} who list a total of 15 candidate VeLLOs as a result of surveying the \textsl{Spitzer} c2d data. The fact that the four above-mentioned VeLLOs are located nearby and at least as distant as the Pipe Nebula means that we can use them to define flux cut-off values at both 24~$\mu$m and 70~$\mu$m to include these currently known VeLLOs. These flux limits ensure that we can detect even the lowest-mass protostellar objects.

As our first selection criterion we thus require candidate YSOs to have flux densities of at least 10~mJy at 24~$\mu$m and 100~mJy at 70~$\mu$m. It turns out that both cut-off values are comfortably above the detection limits in both bands. In the list of sources detected in both MIPS bands, the typical photometric uncertainties determined by MOPEX are 0.09~mJy at the 24~$\mu$m flux cutoff and 5.65~mJy at the 70~$\mu$m flux cutoff. As a second step, we are using a direct application of the original definition of YSO SED classes of \citet{lad87}, although implemented in longer-wavelength (MIPS) bands. Thus, we require a positive spectral index, $\alpha = dlog\lambda F_\lambda/dlog\lambda$, in the two MIPS bands, in order to find the earliest (class I) protostellar stages. In this case, a positive spectral index in $\lambda F_\lambda$ corresponds to a flux ratio of $F_\nu(70\,\mu \rm m)/F_\nu(24\,\mu \rm m)> 70/24$. A similar criterion can be used to select candidate class~II sources, dominated by their disk emission. They are expected to have flux ratios of $F_\nu(70\,\mu \rm m)/F_\nu(24\,\mu \rm m)> 1/3$, corresponding to a spectral index of $\alpha>-2$. Sources above our flux cut-off values but below this minimum spectral index include candidate class~III sources. Since these presumably become difficult to distinguish from stellar photospheres of field stars, we do not discuss these in further detail, focusing only on candidate class~I and II sources as possible products of recent star formation. Only these sources, as defined by our MIPS-based criteria, will be referred to in the following as candidate YSOs. 

\begin{figure*}
\centering
\includegraphics[width=\linewidth]{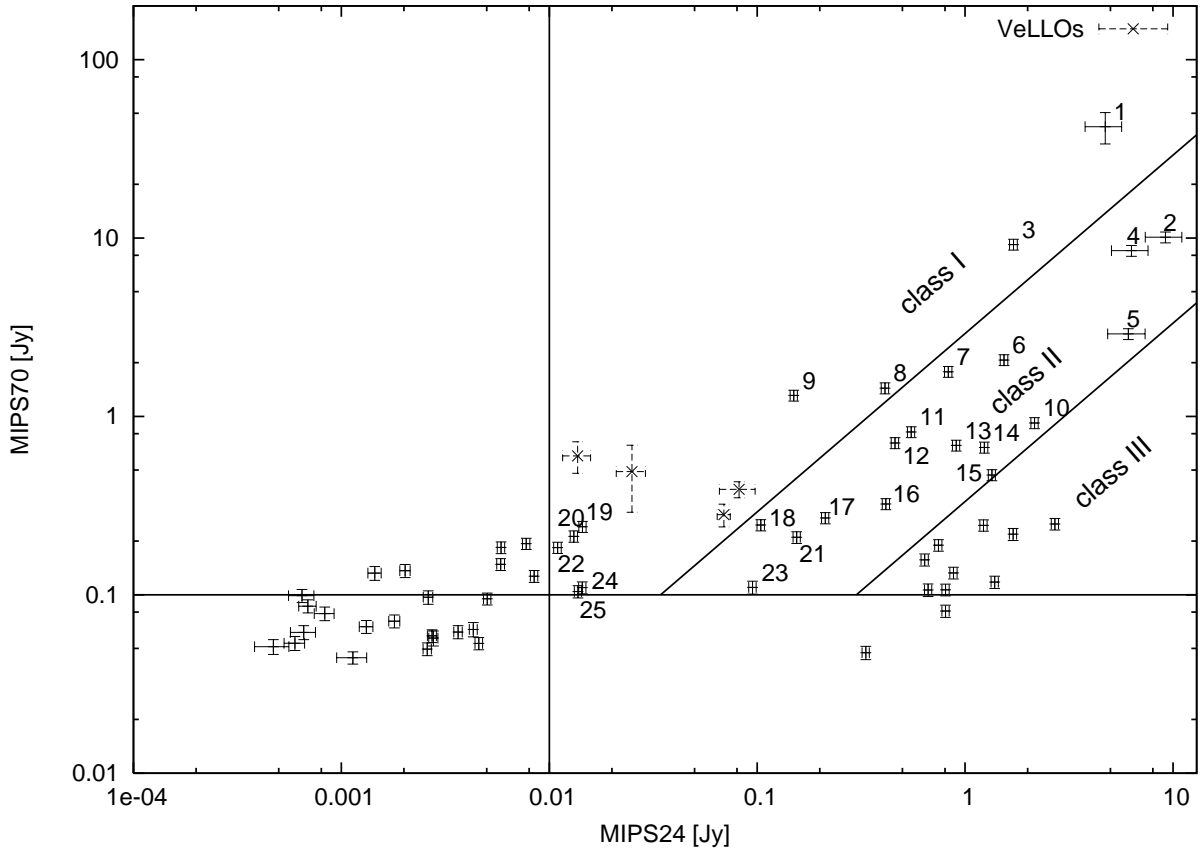}
\caption{Plot of the MIPS fluxes of sources in the high-exinction regions. Regions where we expect to find candidate Young Stellar Objects in various evolutionary stages are marked. The boundaries of these regions were adapted from the original definition of YSO classes at shorter wavelengths, as discussed in Section\,\ref{sec_selec}. Dashed error bars denote the four VeLLOs that have been studied in detail (see text).\label{fig_selyso}}
\end{figure*}

Using these criteria, we identify in Fig.~\ref{fig_selyso} a total of only 25 candidate YSOs in the high-extinction regions of the entire complex, 12 of which are located in B\,59. These sources were matched with the 2MASS point source catalog, using a search radius of 3$''$. The median difference in position is 1.5$''$.
We found one additional candidate YSO in a region without coverage at 70~$\mu$m. This source is listed as candidate~26. Lacking 70~$\mu$m coverage, source 26 was identified by its $K_S-[24]$ color of 10.459$\pm$0.082~mag. There are no further sources with comparably extreme $K_S-[24]$ colors that are not already covered by the above criteria. Given the superior sensitivity of the above method, which does not depend on near-infrared detections, we do not further discuss $K_S-[24]$ colors.
The resulting 26~candidate YSOs and their photometry are listed in Tables~\ref{tab_candsel} and \ref{tab_candsel2}, sorted by descending 70$\mu$m flux. The photometric errors consist of the fitting errors reported by MOPEX, with the absolute errors\footnote{4\% at 24~$\mu$m and 7\% at 70~$\mu$m, according to MIPS Data Handbook V3.3.1} added in quadrature (for the saturated sources, a larger error is assumed). In order to allow us to compare the photometry, we have added these absolute uncertainties also to the errors reported by \citet{bro07} for the non-saturated sources, since the errors in their source table are the formal fitting errors. For all candidate YSOs we confirmed their detection in two different epochs of MIPS24 observations, ensuring that they were not caused by asteroids. Source 26 was confirmed by the presence of a mid-infrared IRAC counterpart.

For B\,59, when applying our criteria to the 12 sources from \citet{bro07} that have MIPS24 and MIPS70 photometry, we find four candidate class~I sources and eight candidate class~II sources. This leaves eight sources from the \citet{bro07} list of 20 that do not appear to qualify as candidate YSOs according to our MIPS-based criteria. While in some cases, these objects are genuinely excluded by our method, in other cases, no suitable MIPS photometry is available. One of these sources, [BHB2007] 20 is not covered at 70~$\mu$m so that it is not selected by our criteria. Three sources ([BHB2007] 2, 6, and 12) are too close to bright neighboring sources for any reliable photometry in either of the MIPS bands to be extracted (and no such photometry is listed in \citealp{bro07}); therefore, these sources do not appear in our list. The remaining four sources are the outermost objects which are not detected at 70~$\mu$m; only upper limits were given by \citet{bro07}. These include [BHB2007] 5 and 17, which Covey et al. (2009) determined to be background giants, using near-infrared spectroscopy. The source [BHB2007]~17 is very close to the chip edge in the 70~$\mu$m coverage, and not detected in spite of its bright 24~$\mu$m counterpart. 

The sensitivity of our MIPS mosaics, as we have argued, is enough to even detect VeLLOs at the distance of the Pipe Nebula. Therefore, we estimate that our survey is complete for candidate class~I protostars of $L>0.2$\,$L_\odot$. We note that even for candidate class~II sources, our selection method is very sensitive. Note that we used a conservative limit of $\alpha_{24-70}>-2$ to \textsl{find} all possible candidate class~II sources. This index corresponds to the ideal case of an optically thick, spatially flat disk. In reality, observed class~II SEDs often have significantly flatter slopes due to disk flaring and viscuous heating, resulting in $\alpha = -0.7\pm0.3$ \citep[][for a spectral range of 2 -- 60\,$\mu$m]{ken87}. In Section~\ref{closerlook}, when we study their SEDs, we will see from the final classification (Table\,\ref{tab_candfit}) that our candidate class~II sources have a median\footnote{The values range from $\alpha_{K-24}=-0.54$ to $\alpha_{K-24}=-0.83$, leaving aside source 24 with a markedly double-peaked SED ($\alpha_{K-24}=-1.47$).} spectral index of $\alpha_{K-24}=-0.73$. We can thus roughly estimate the completeness for such typical class~II sources by comparing with an observed example, the fluxes of FM~Tau, a K3 class~II source ($\alpha_{K-24}=-0.65$) in Taurus at a similar distance as the Pipe. FM~Tau has a total luminosity of 0.5~$L_\odot$ (e.g., \citealp{str88}) and a 70~$\mu$m flux density of 0.29~Jy (from the Taurus Spitzer Legacy Project). Scaling that to our flux density cut-off at 70~$\mu$m (0.1~Jy) and assuming the same distance for the sake of this comparison yields a completeness limit for class~II SEDs of only $L_{\rm bol}\sim0.2$\,$L_\odot$. The flux density ratio $F_\nu(70\,\mu \rm m)/F_\nu(24\,\mu \rm m)$ in this case is 0.59. A similar check with two late-type class~II objects, FN Tau (M5, $\alpha_{K-24}=-0.53$) and FP Tau (M2.5, $\alpha_{K-24}=-1.12$), yields even lower limiting luminosities for our flux limits (i.e., 0.05\,$L_\odot$ and 0.12\,$L_\odot$, respectively), corroborating our estimate of $L_{\rm bol}\sim0.2$\,$L_\odot$. The completeness level for typical observed class II sources is comparable to the one for class~I protostars discussed above.

While all candidate sources by definition are in regions with $A_V>1.2$~mag (in the extinction map without extended structure, see above), not all of them are projected onto what \citet{rat09} define as dense cores, see Table~\ref{tab_candsel}. Five of our sources either lie in extinction cores that do not fulfill the minimum size requirement set by \citet{rat09} or were excluded due to their C$^{18}$O emission properties: One source is close to an extinction peak without detected C$^{18}$O emission (source 16), and, most interestingly, the extinction core onto which source 17 is projected, as well as its neigboring cores, have C$^{18}$O emission at a velocity that is significantly different from the cloud velocity ($v=21$\,km\,s$^{-1}$ vs. a range for the Pipe Nebula of $v=2..8$\,km\,s$^{-1}$, \citealp{rat09}). This candidate YSO therefore apparently does not belong to the Pipe Nebula complex. For the five cases where cores and their masses are not listed by \citet{rat09}, Table~\ref{tab_candsel} lists corresponding mass estimates, derived according to the procedure outlined in Section~\ref{sect_sfe}.  These estimates take into account all pixels of the background-subtracted extinction map with $A_V>1.2$~mag if they have not been assigned to neighboring cores. Note that some of the changes in the refined core list are due to the intricacies of defining the core population, a process that is not unambiguous.

\subsection{Comparison to control regions}

\begin{figure*}
\centering
\includegraphics[width=\linewidth]{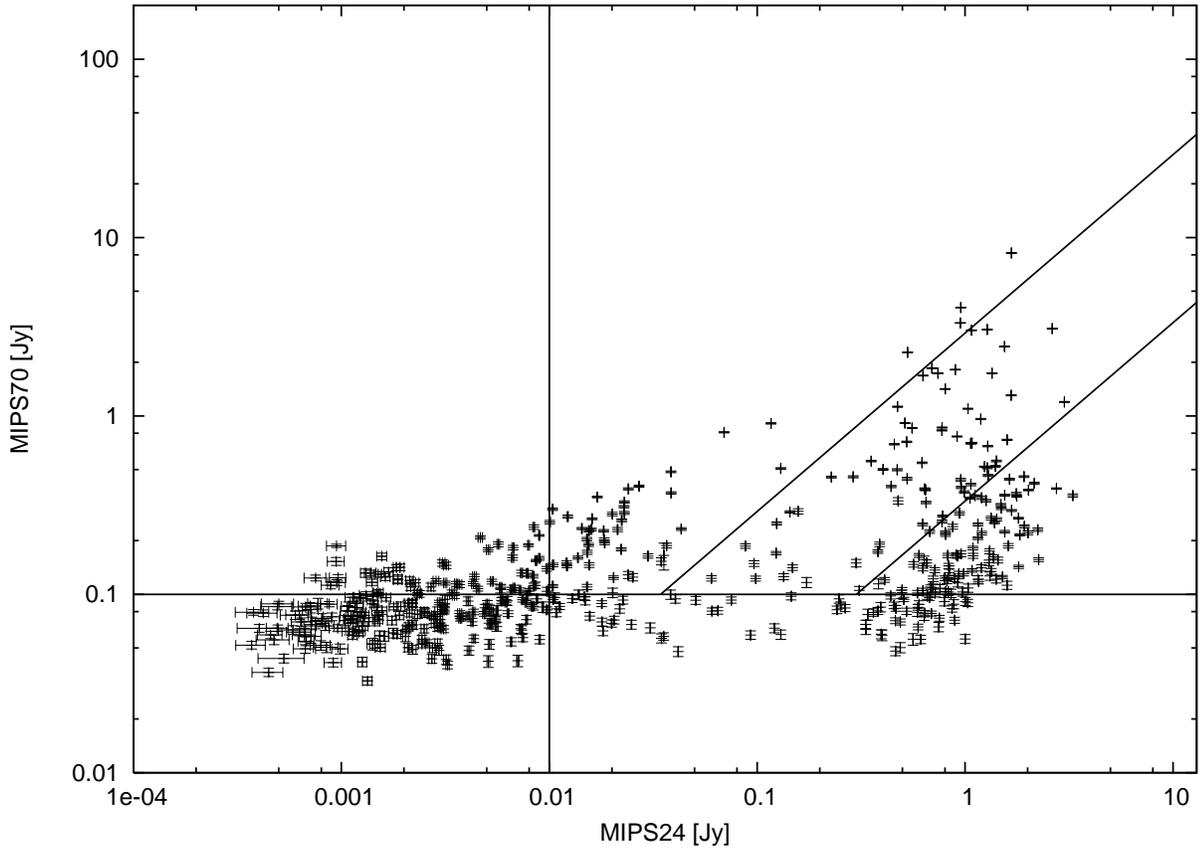}
\caption{Plot of the MIPS fluxes of sources in the control region of the maps, outside of the high-extinction regions ($A_V<1.2$). The areas used for the identification of candidate Young Stellar Objects are marked, as in Fig.~\ref{fig_selyso}, even though we only use this plot to determine the background contamination.
\label{fig_selbkg}}
\end{figure*}

While we have estimated the sensitivity of our selection method, we still have to determine the amount of contamination by background sources. As control region, we use those parts of our MIPS mosaics that are not covering the high-extinction regions, i.e., regions with $A_V<1.2$. Applying the selection criteria to these sources allows us to approximately predict how many detections in each category could be due to background objects and to judge whether the number of identified candidates exceeds the number of expected background sources. As mentioned above, about 94\% of our map is not covering high-extinction regions. Given the proximity of the Pipe Nebula to the Galactic center, there is a high density of background sources, presumably mainly in the Galactic bulge. While 56 sources were detected at both 24~$\mu$m and 70~$\mu$m in regions with extinctions above 1.2~mag (not applying any filtering), 490 such sources were detected in the remainder of the map. Fig.~\ref{fig_selbkg} depicts the sources in the control region in the same way as the high-extinction sources were plotted in Fig.~\ref{fig_selyso}. Some of the sources in the control region technically fulfill our candidate YSO selection criteria, indicating that there is contamination in our list of candidate YSOs. Table~\ref{tab_statcomp} contains a comparison of the control region and the high-extinction regions in the Pipe Nebula in terms of candidate YSOs. Taking into account that regions with extinctions above 1.2~mag cover only $\sim7$\% of the map (see above), we find that that in the three ranges of flux ratios corresponding to YSOs, we find more candidates than can be expected from background contamination alone. In particular, and most importantly, we detect nine candidate class~I sources while the expected background contamination is 2.9~sources. This relative overdensity of candidate YSOs in the high-extinction regions suggests that their occurence is indeed related to the presence of dense material and supports the notion that some of these objects really are YSOs.

\subsection{Comparison to other star-forming regions}
\label{other.SF}

\begin{figure*}
\centering
\includegraphics[width=0.9\linewidth]{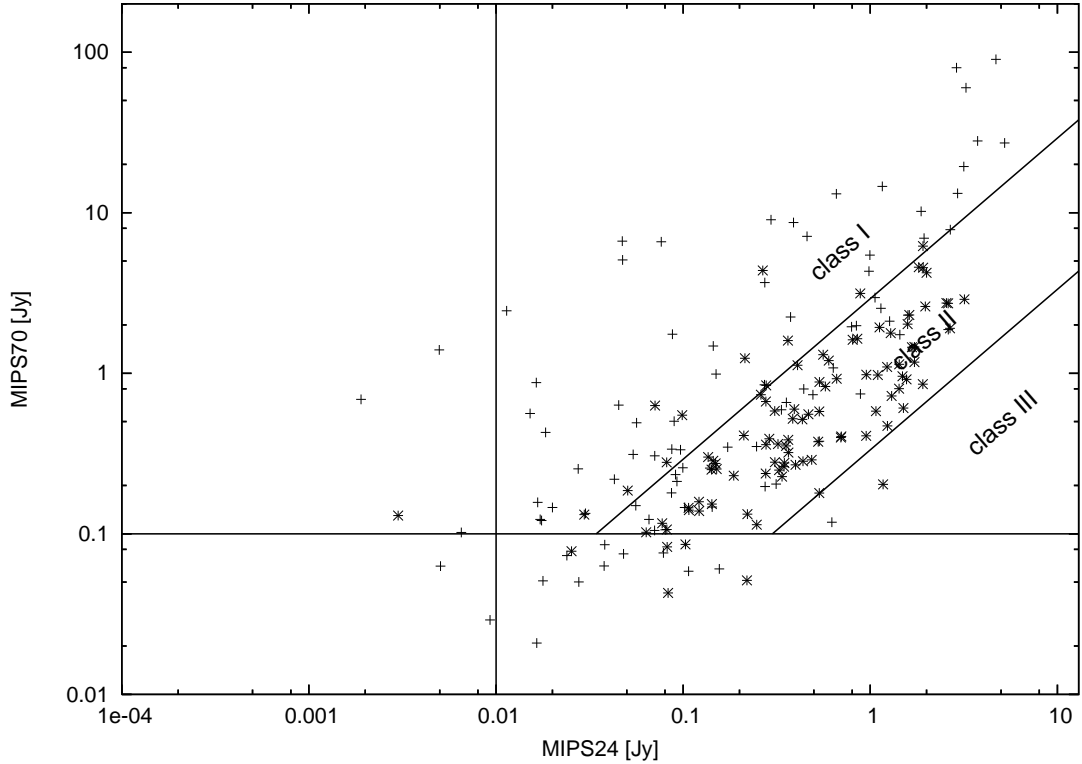}
\caption{Plot of the MIPS sources in other star forming regions: `+' symbols mark MIPS sources among the candidate YSOs from the ensemble list of cores produced by the Cores to disks Spitzer Legacy Project. The lines denote the same selection regions that we also used for the Pipe candidate YSOs. Note that the plot contains objects from star-forming regions at different distances, without a correction for this effect. Asterisks mark the MIPS fluxes of counterparts to YSOs in Taurus-Auriga, as listed by \citet{ken08}, see text. For comparison, the flux regions for our YSO selection are shown. Both the c2d sources as well as the Taurus sources mostly fall into the area that we use for candidate YSO identification. \label{fig_c2d_taurus}}
\end{figure*}

There are two obvious possibilities to compare the application of our candidate YSO identification method to other star-forming regions. First of all, given its location at about the same distance, it is interesting to compare the Pipe Nebula to the Taurus star-forming region. While a full application of our method to the Taurus data is beyond the scope of this paper, we can at least check whether known objects occupy the region of flux space that we use to identify candidate YSOs. In a preliminary version of their source list, the Taurus Spitzer legacy project\footnote{http://data.spitzer.caltech.edu/popular/taurus2/ 20081007\_enhanced/} lists 693 sources that are detected at both 24~$\mu$m and 70~$\mu$m, a pre-condition for our candidate YSO selection method. The Taurus map covers a total of about 40~square degrees, four times larger than our map of the Pipe Nebula. We correlate the \textsl{Spitzer} source list with the list of YSOs in Taurus-Auriga given by \citet{ken08}. This table contains 383 low-mass YSOs of which 98 have counterparts in the \textsl{Spitzer} list with detections in both MIPS bands (within a search radius of 1$''$, see above; note that the two lists cover different areas). The Kenyon et al. list does not distinguish the different evolutionary stages, but as shown in Fig.~\ref{fig_c2d_taurus}, this selection of sources is well constrained by the regions in the 70~$\mu$m vs. 24~$\mu$m plot that we use for the selection of candidate YSOs. To illustrate that prototypical class~I sources fall into the corresponding selection region, we note that L1551~IRS5 (not covered by the Taurus Spitzer Legacy Project), already based on its IRAS data, fulfills the criteria, but lies outside of our plot limits due to its brightness: its 60$\mu$m flux is 373~Jy, and its 25$\mu$m flux is 106~Jy, i.e., the flux ratio is 3.5. This example once more underlines the sensitivity of the \textsl{Spitzer} MIPS observations.

For a second comparison, we consider the candidate YSOs that have been identified by the ``Cores to disks'' (c2d) Spitzer Legacy Project. This comparison is different, since it allows us to select already identified candidate YSOs and look at their properties at 24~$\mu$m and 70~$\mu$m. The c2d ``Ensemble Cores'' list of candidate YSOs contains 106 objects with both 24~$\mu$m and 70~$\mu$m fluxes out of 122 candidate YSOs that were observed at both wavelengths\footnote{Sixteen sources are listed as ``undetected'' at 70~$\mu$m, but are located in the vicinity of other, mostly bright, 70~$\mu$m sources, complicating their detection.}. Reassuringly, when we plot the sources in the same way as we plotted the candidate YSOs in the Pipe, the sources roughly fall into the region of flux space where we expect candidate YSOs in the Pipe cloud (Fig.~\ref{fig_c2d_taurus}). This is reassuring since the MIPS spectral index is not part of the selection criteria that were used by the c2d team. Note that the data have not been corrected for the different distances ($\sim120$--250~pc) of the several different nearby star-forming regions that were part of the survey.

These two examples demonstrate that using the 24~$\mu$m and 70~$\mu$m MIPS bands and their high sensitivity constitutes a powerful method to look for candidate YSOs in the Pipe cloud. The drawback of relatively low angular resolution is in our case overcompensated by the enormous reduction of sources to survey (while keeping a high sensitivity) compared to the use of IRAC data which in this direction of the Galaxy are affected by source crowding.

\subsection{A closer look at the candidate YSOs}
\label{closerlook}

A cross-correlation of the 26 candidate YSOs with catalogs, using SIMBAD and Vizier, allows us to identify a first subset of sources that can be rejected as YSO candidates. Source~10 is a 1.4~GHz radio continuum source in the NRAO VLA Sky Survey and listed as a possible pre-planetary nebula by \citet{koh01}. Sources 13, 14, and 15 are all listed as high mass-loss asymptotic giant branch stars by \citet{ojh07}. After a comparison with typical \textsl{Spitzer} galaxy SEDs, in particular looking at the influence of PAH features on the IRAC photometry \citep[e.g., ][]{row05}, three additional sources, namely 19, 20, and 22 have been rejected as being likely galaxies. We note that their IRAC colors are [5.8$\mu$m]-[8.0$\mu$m]$>$2 and [3.6$\mu$m]-[4.5$\mu$m]$<$0.9, supporting this conclusion \citep[e.g., ][]{rob08}. Source~25 looks photospheric except for its 70~$\mu$m flux which is the weakest in this selection. In fact, the 70~$\mu$m detection is surrounded by weak extended emission leading to an overestimate of the flux in the PRF fit. The source has been rejected as not being a true excess source; it is clearly not a class I or class II YSO. The use of catalogs also allows us to confirm the selection of a YSO: source 5 is the Herbig AeBe star KK~Oph which is very likely associated to B\,59; it appears to have a T~Tau companion \citep{her05,car07}. Additionally, Covey et al. (2009) have analyzed near-infrared spectroscopy of our sources 11 and 16, confirming only the latter as a YSO (their source 2, a binary system unresolved in our data), while leaving open the identification of the former (source 1 in their designation). \citet{ria09} confirm the protostellar nature of source 1. Thus, after this first cross-correlation with catalogs and previously obtained data as well as comparison to galaxy SEDs, 18 candidate YSOs remain, only six of them outside of B\,59. We note that the two \citet{bro07} sources rejected as giant stars by Covey et al. (2009), [BHB2007] 5 and 17, are not detected at 70~$\mu$m and thus did not appear in our list. 

\begin{figure*}
\centering
\includegraphics[width=0.9\linewidth]{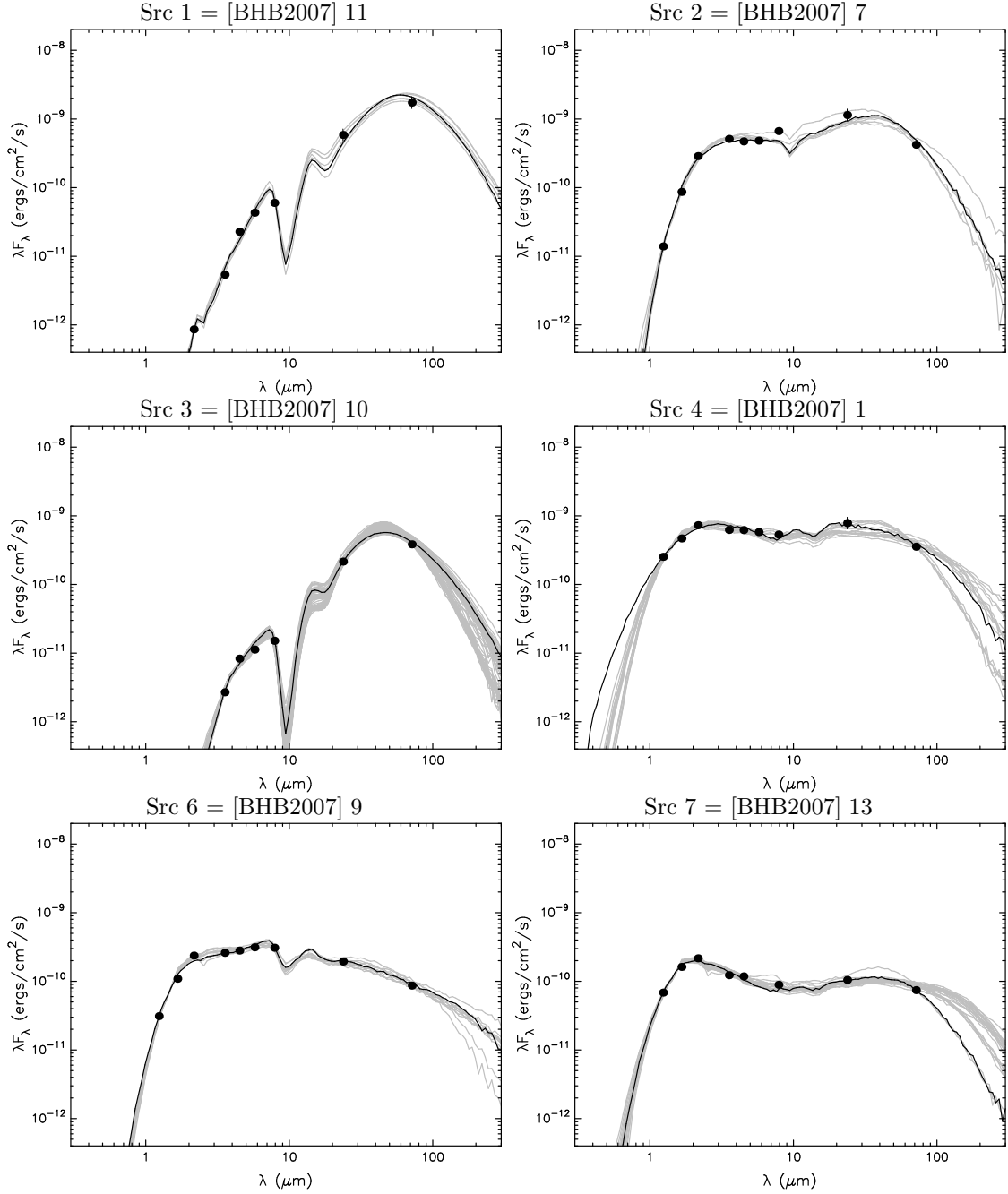}
\caption{SEDs of candidate YSOs in B\,59. The black line shows the best fit with gray lines delineating alternative fits within the goodness-of-fit criterion used (see text). Note that for Source 9, this criterion had to be relaxed. \label{fig_seds1}}
\end{figure*}

\begin{figure*}
\centering
\includegraphics[width=0.9\linewidth]{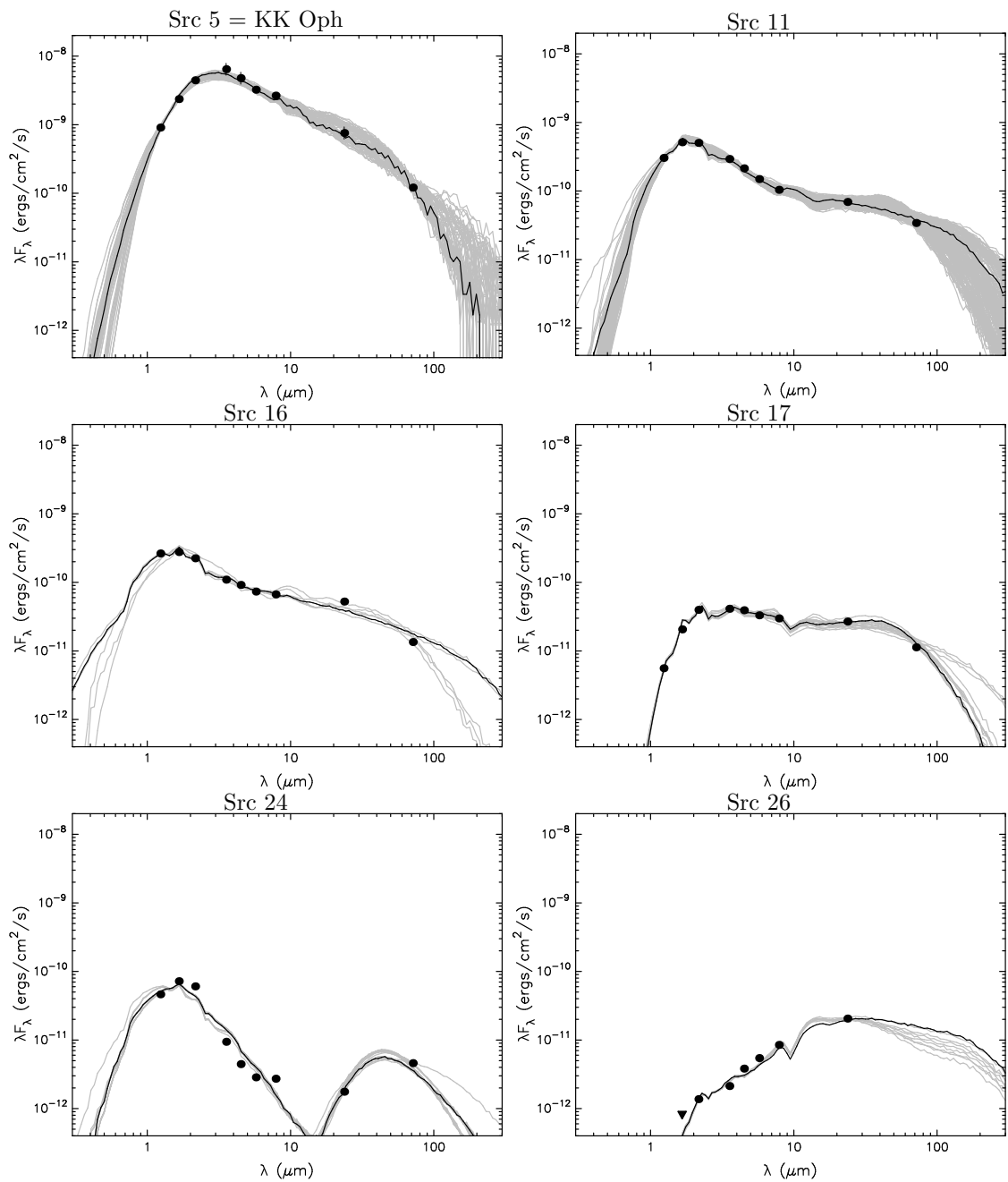}
\caption{SEDs for candidate YSOs outside of B\,59 \label{fig_seds2}}
\end{figure*}

\begin{figure*}
\centering
\includegraphics[width=0.9\linewidth]{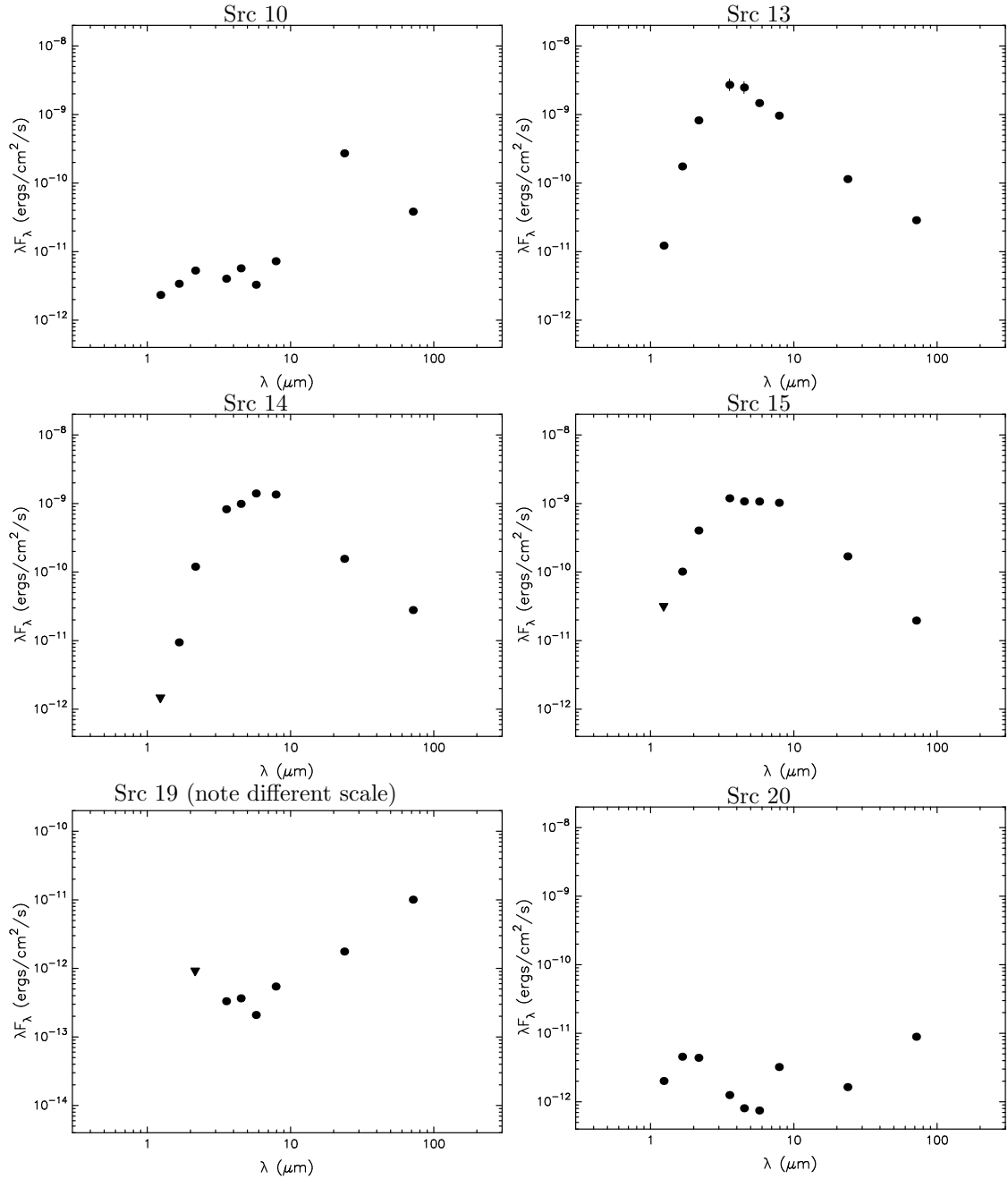}
\caption{SEDs of rejected candidate YSOs and four sample control sources that fulfill the YSO selection criteria (designated only with their coordinates). \label{fig_seds3}}
\end{figure*}

In order to better characterize the candidate YSOs, we have determined their infrared SEDs, based on 2MASS, IRAC and MIPS photometry. For comparison, we have also determined SEDs for some sources in the control regions that would qualify as candidate YSOs.  Subsequently, to test whether the SEDs can be explained as protostellar, we have attempted to fit the SEDs of candidate YSOs with models from \citet{rob07}. These pre-computed radiative-transfer models cover a large region of parameter space. Since most of the sources are relatively bright 2MASS sources and thus do not look like edge-on disk configurations, we excluded such models in the grid from the fitting process, realizing that all SEDs can still be well fitted and explained. For the fits, photometric errors larger than the nominal errors were assumed: 0.1~mag for 2MASS and 10\% for all \textsl{Spitzer} photometry except for data corrected for saturation, where we assume 20\%. The SEDs and fit results are shown in Figs.~\ref{fig_seds1}, \ref{fig_seds2}, and \ref{fig_seds3}. Table~\ref{tab_candfit} lists the range of total extinction and the range of bolometric luminosities for all candidate YSOs. The parameter ranges are simply giving the extreme values for all fits within a relatively strict goodness-of-fit criterion\footnote{For source 9, only a single fit was found with this criterion; we therefore relax the criterion for only this source to $\chi^2-\chi_{\rm best}^2<2n_{\rm data}$.}, $\chi^2-\chi_{\rm best}^2<n_{\rm data}$. The fit results show that the candidate YSOs can be reasonably described as a mostly low-mass YSO population. The lowest-luminosity candidate class~I protostar, source 26, might be a VeLLO in its own right. Also, the lowest luminosities deduced from the model fits are of the same order as our estimated completeness limit.

We use the full SEDs and the model fits to reexamine and refine our initial classification of the candidate YSOs which was based solely on the MIPS spectral index. In particular, Figs.~\ref{fig_seds1} and \ref{fig_seds2} show that two candidate YSOs, sources 9 and 24, initially identified as candidate class~I sources, exhibit complex double-peaked SEDs that cannot be adequately described by a single spectral index. The full SEDs show, however, that these two sources are more aptly described as class~II objects. Additionally, an examination of the SEDs of sources initially selected as candidate class~II objects shows that several of these have SEDs that are flatter than expected for prototypical class II objects with shapes similar to those of the flat-spectrum protostellar class defined by \citet{gre94}. Since their SEDs are well-behaved, we use the wide $K$--24\,$\mu$m spectral index, as described in a footnote to Table\,\ref{tab_candsel2}, to refine our classification to include flat-spectrum protostars. The refined YSO classifications are listed in Table\,\ref{tab_candsel2}.

In an attempt to characterize the 48 control sources that technically fulfill our criteria for being candidate class~I protostars, we used SIMBAD to identify counterparts. Such counterparts are found for only nine sources within a search radius of 4$''$. Eight of these appear to be planetary nebulae and one is an OH/IR star. These 48 sources have a mean extinction of 0.1~mag in the extinction map where extended structure has been removed, and a mean extinction of 4.0~mag in the full extinction map. A sample of control SEDs is shown in Fig.~\ref{fig_seds3}. While many sources remain unidentified and the multi-wavelength coverage is not the same for all of these objects, most of the sampled sources appear to be galaxies based on their SED shapes, following the same arguments also used in rejecting candidate YSOs above (mainly the influence of PAH features on the IRAC photometry).

\subsection{Location of the candidate YSOs relative to extinction peaks}

\begin{figure*}
\centering
\includegraphics[width=0.8\linewidth]{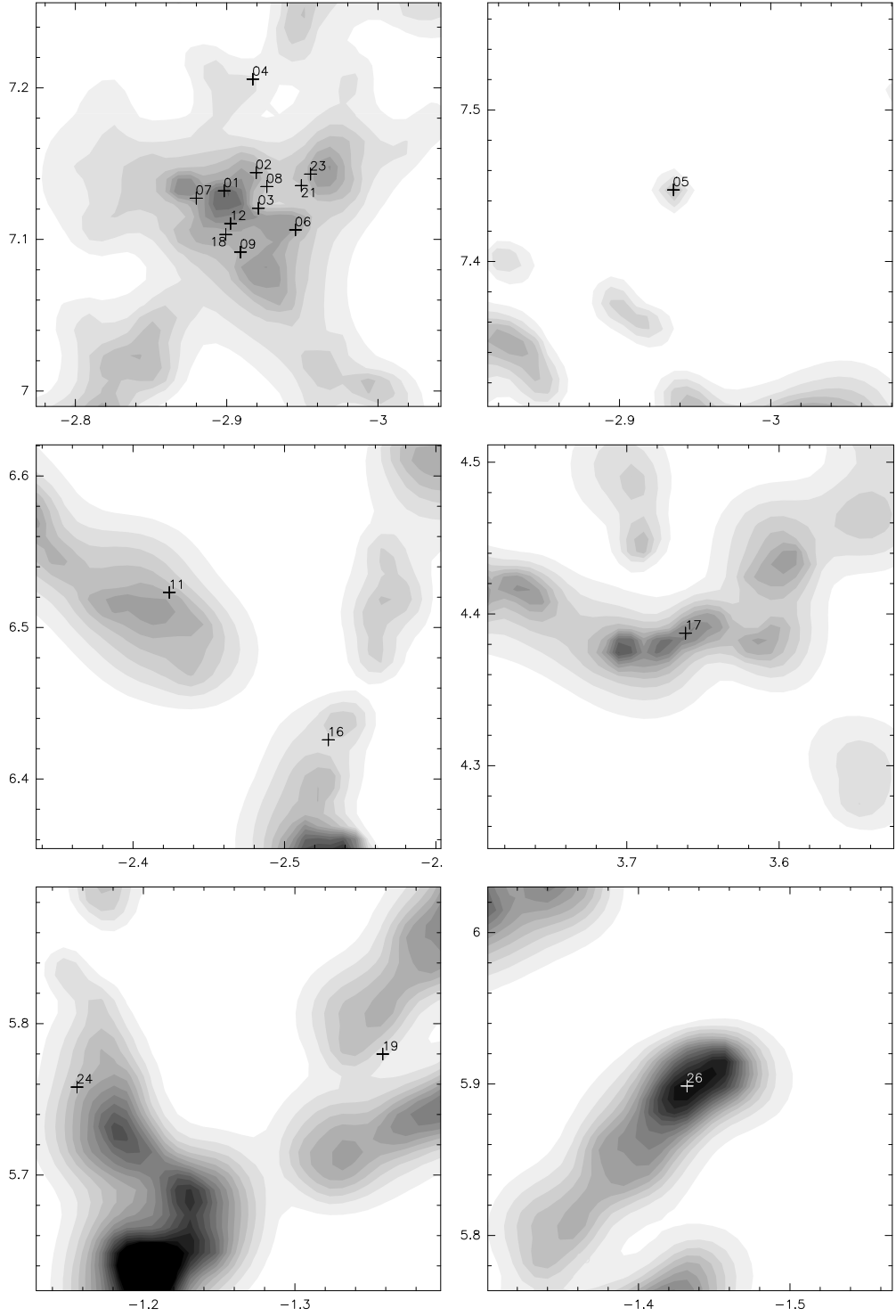}
\caption{Location of the identified candidate YSOs relative to extinction peaks. Extinction is shown in shades of grey, starting at $A_V<1.2$ and advancing in steps of 0.5. Due to its high extinction, the B\,59 region is instead shown in an $A_V$ range from 2 to 20 in steps of 2. The coordinates are Galactic coordinates.\label{fig_extmapviews}}
\end{figure*}

\begin{figure*}
\centering
\includegraphics[width=\linewidth]{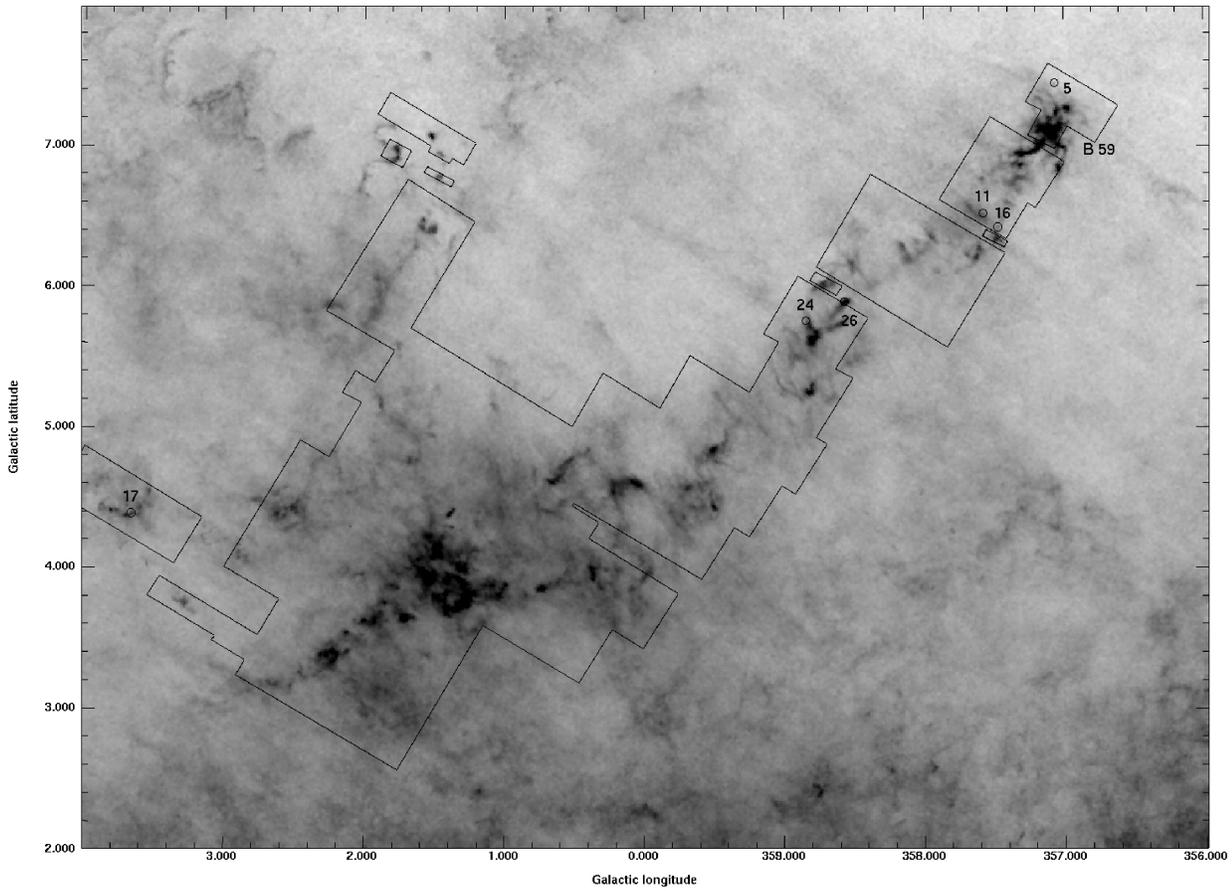}
\caption{Location of the identified candidate YSOs on the full extinction map from \citet{lom06}. Due to the high source density in B\,59, only the location of the cluster is indicated. The approximate outline of the combined \textsl{Spitzer}-MIPS coverage (24\,$\mu$m and 70\,$\mu$m) is marked. For details, see Fig.~\ref{fig_extmapviews}. \label{extmap_full}}
\end{figure*}

Very young candidate YSOs would most likely be found at or in close proximity of the extinction peaks. Fig.~\ref{fig_extmapviews} shows the identified candidate YSOs overlaid onto the corresponding sections of the extinction map. Given that the candidate YSOs were selected for being inside areas of the extinction map that are above a minimum threshold, it is not surprising to see them close to the extinction peaks, but few are centered on peaks, the best case being source 26. Since one would expect the youngest sources (class I) to be located at or very near to the extinction peaks, the observed offsets indicate that the candidate YSOs are likely predominantly in later evolutionary stages (i.e., class~II). This is compatible with our flux-based characterization which found that most of the candidate YSOs are class~II rather than class~I (see above). It is interesting to note that while five of the six candidate YSOs outside of B\,59 are within 2.3$^\circ$ of the cluster, a single candidate (source 17) is 5.9$^\circ$ away, at the ``opposite end'' of the cloud. We note again that the corresponding core as well as the surrounding cores were rejected by \citet{rat09} as being part of the Pipe complex since their C$^{18}$O velocities are significantly offset from the system velocity. While source 17 nevertheless is a candidate YSO, it probably does not belong to the Pipe Nebula complex.
Fig.~\ref{extmap_full} shows the six candidate YSOs outside of B\,59 on the extinction map from \citet{lom06}. It is indeed interesting to note that most of them are located in the ``stem'' of the Pipe, none having been detected in the ``bowl'', even though there are regions of considerable extinction in that area. This suggests that the area around B\,59 at the tip of the stem is the most evolved region, followed by the stem, leaving the entire ``bowl'' in a very early evolutionary stage. Interestingly, this is in line with the conclusions of \citet{alf08} who presented an optical polarization study of the Pipe Nebula. From a strong polarization gradient across the Pipe Nebula, with the lowest amounts of polarization found towards B\,59, they propose a very similar picture. While they also suggest that the entire ``bowl'' region is in a very early stage, they conclude that the ``stem'' is about to form stars while B\,59 has actively ongoing star formation.

\subsection{Star formation efficiency in the Pipe Nebula}
\label{sect_sfe}

In total, we have found 21 candidate YSOs in the Pipe Nebula. Fifteen\footnote{In addition to the 12 MIPS-selected candidate YSOs, we consider the three above-mentioned \citet{bro07} sources in B\,59 for which no photometry could be obtained due to nearby bright neighbors, yielding an upper limit of 15 YSOs in B\,59.} of these are in B\,59 and only six in the entire remaining cloud complex. Star formation efficiencies (SFEs) are usually given as SFE$=M_*/(M_*+M_{\rm cloud})$, i.e., they are compared to the total mass of the cloud (in this case, $M\sim10^4\,M_\odot$, \citealp{oni99}). Assuming stellar masses of $\sim0.3\,M_\odot$ each, this translates into an extremely low star formation efficiency when compared to the mass of the entire cloud, it is only $\sim0.06$\%. Typical values are a few percent; the median of 2\% given by \citet{mye86} is 33 times higher than our value for the Pipe Nebula. 

Furthermore, we can estimate the star formation efficiency for the densest material, considering that the high-extinction regions represent the dense component of the cloud. We can estimate the corresponding cloud mass from the summed pixels in the extinction map, taking into account the high-extinction regions with $A_V>1.2$~mag (in the background-corrected map) that were surveyed for candidate YSOs:

\begin{equation}
M = \mu {\rm m_{H}} \beta A_{\rm pix} \Sigma A_V
\end{equation}

Here, $\mu$ is the mean molecular weight (1.37), ${\rm m_{H}}$ is the mass of the hydrogen atom, $\beta$ is the conversion from optical extinction to column density, and $\Sigma A_V$ is the summed optical extinction. $A_{\rm pix}$ is the area of a pixel in cm$^{-2}$ at the distance of the source, for extinction pixels with a side length of $s_{\rm pix}$ [$''$]: $A_{\rm pix} = (\pi/180/3600)^2 s_{\rm pix}^2 d^2$, where $d$ is the distance in cm. While there are 14066 pixels in the extinction map above a threshold of $A_V>1.2$~mag (each of size $30''\times30''$), only 11063 of these are covered by both the 24~$\mu$m and the 70~$\mu$m maps. The summed $A_V$ in these pixels is 29751~mag. Using a conversion factor of $\beta=2\times 10^{21}$~cm$^{-2}$mag$^{-1}$ (e.g., \citealp{boh78}), the resulting mass is $230\,M_\odot$. Assuming an upper limit of 21 candidate YSOs, each approximately of $\sim0.3\,M_\odot$, again yields an extremely low star formation efficiency of 2.7\% when compared to the densest material. This is considerably lower than typical values for clusters which are on the order of 10--40\% \citep[e.g., ][]{lad91}. We note that at most five of the identified 134 extinction cores (or 3.7\%) have candidate YSOs, not counting three additional candidate YSOs in extinction cores projected onto high-extinction regions not part of cores as defined by \citet{rat09}.
For the combined data of the \textsl{Spitzer} c2d project, the average star formation efficiency is 4.8\% while it is much higher when compared to the dense material, as derived from dust continuum observations \citep{eva09}.

\section{Summary}

We have used very sensitive mid-infrared mapping of the Pipe Nebula to obtain a sensitive census of its extremely low star formation activity. Using the 24~$\mu$m and 70~$\mu$m MIPS bands to identify candidate YSOs is a powerful tool, allowing us to search the cloud for protostars down to objects with very low luminosities. For both class~I and typical class~II YSOs the search is estimated to be complete down to luminosities of $\sim0.2$\,$L_\odot$. The star formation efficiency in the Pipe Nebula is only $\sim0.06$\,\% when compared to the total cloud mass and 2.7\% when compared to the mass of the high-extinction regions. This is more than an order of magnitude lower than in typical star-forming regions. Candidate YSOs are found in at most five of the 134 identified extinction cores, leaving the overwhelming majority of cores apparently starless. Outside of Barnard~59, where a previously known cluster of YSOs resides, we find only six new candidate YSOs in the entire remaining Pipe Nebula, one of which has already been confirmed as a YSO by near-infrared spectroscopy (Covey et al. 2009, in prep.). One of these candidate YSOs very likely is not a member of the Pipe Nebula complex. Using our MIPS-based selection criteria, we find an upper limit of 15 candidate YSOs in B\,59, all of them previously known. Interestingly, the Pipe candidate YSOs outside of B\,59 are not evenly distributed over the remaining cloud, but they are within 2.3$^\circ$ of B\,59, leaving large regions of the cloud, particularly the bowl region, without any candidate YSOs. These results clearly demonstrate that apart from B\,59, the Pipe Nebula is a region with extremely low star formation activity. We conclude that it appears to be an ideal region to study the initial conditions and earliest stages of star formation.

\acknowledgments{We would like to thank Thomas Robitaille for help with the SED fits. This work is based on observations made with the Spitzer Space Telescope, which is operated by the Jet Propulsion Laboratory, California Institute of Technology under a contract with NASA. Support for this work was provided by NASA through an award issued by JPL/Caltech, contract 1279166. This publication makes use of data products from the Two Micron All Sky Survey, which is a joint project of the University of Massachusetts and the Infrared Processing and Analysis Center/California Institute of Technology, funded by the National Aeronautics and Space Administration and the National Science Foundation.}

{\it Facilities:} \facility{SST (MIPS), SST (IRAC)}

\begin{deluxetable}{rrrrrrrrrrrrr}
\tabletypesize{\scriptsize}
\rotate
\tablecaption{Candidate young stellar objects in the Pipe Nebula\label{tab_candsel}}
\tablewidth{0pt}
\tablecolumns{13}
\tablehead{
\colhead{no.} & \colhead{RA, Dec (J2000)\tablenotemark{a}} & \colhead{core} & \colhead{core mass} & \colhead{$A_V$\tablenotemark{b}} & \colhead{$A_V$\tablenotemark{c}} & \colhead{MIPS70} &
\colhead{err.} & \colhead{$\Delta$pos\tablenotemark{d}} & \colhead{MIPS24} & \colhead{err.} & \colhead{$F_\nu(70\,\mu \rm m)/$} & \colhead{ID\tablenotemark{e}}\\
\colhead{} & \colhead{} & \colhead{} & \colhead{[$M_\odot$]} & \colhead{[mag]} & \colhead{[mag]} & \colhead{[Jy]} & \colhead{[Jy]} & \colhead{[$''$]} & \colhead{[Jy]} & \colhead{[Jy]} & \colhead{$F_\nu(24\,\mu \rm m)$} & \colhead{}
}
\startdata
1  & 17:11:23.09 --27:24:32.80 & 9   & 19.4 & 20.2 & 24.9 & 42.100 & 8.400   & 0.5 &  4.715 & 0.940   &  8.9 & [BHB2007] 11    \\ 
2  & 17:11:17.45 --27:25:08.48 & 9   & 19.4 & 11.1 & 15.0 & 10.100 & 0.709   & 3.6 &  9.200 & 1.840   &  1.1 & [BHB2007] 7     \\ 
3  & 17:11:22.12 --27:26:02.30 & 9   & 19.4 &  8.8 &  9.9 &  9.190 & 0.644   & 0.7 &  1.710 & 0.074   &  5.4 & [BHB2007] 10    \\ 
4  & 17:11:03.95 --27:22:55.39 & 7   &  4.0 &  4.2 &  7.9 &  8.490 & 0.596   & 0.6 &  6.320 & 1.260   &  1.3 & [BHB2007] 1\tablenotemark{f}\\
5  & 17:10:08.20 --27:15:20.20 & --  &$<$0.1&  2.4 &  3.0 &  2.901 & 0.203   & 1.5 &  6.100 & 1.250   &  0.5 & KK Oph   \\
6  & 17:11:21.43 --27:27:42.91 & 9   & 19.4 & 12.4 & 15.7 &  2.070 & 0.145   & 1.4 &  1.540 & 0.064   &  1.3 & [BHB2007] 9	\\
7  & 17:11:27.00 --27:23:49.05 & 9   & 19.4 & 10.5 & 18.1 &  1.780 & 0.125   & 0.6 &  0.831 & 0.034   &  2.1 & [BHB2007] 13	\\
8  & 17:11:18.13 --27:25:49.30 & 9   & 19.4 &  7.5 & 13.0 &  1.440 & 0.101   & --  &  0.412 & 0.017   &  3.5 & [BHB2007] 8     \\
9  & 17:11:30.32 --27:26:29.57 & 9   & 19.4 & 10.3 & 14.4 &  1.310 & 0.092   & 0.9 &  0.150 & 0.006   &  8.7 & [BHB2007] 16	\\
10 & 17:35:02.22 --25:23:49.54 & --  &  0.3 &  2.2 &  8.5 &  0.919 & 0.065   & 0.6 &  2.160 & 0.086   &  0.4 & $R$	\\
11 & 17:14:51.05 --27:22:41.02 & 17  &  1.6 &  2.2 &  5.6 &  0.817 & 0.058   & 0.8 &  0.550 & 0.022   &  1.5 & Src 1\tablenotemark{g}	\\
12 & 17:11:27.28 --27:25:28.99 & 9   & 19.4 & 12.4 & 20.3 &  0.710 & 0.050   & 1.5 &  0.455 & 0.018   &  1.6 & [BHB2007] 14	\\
13 & 17:29:41.64 --26:29:09.34 & 49  &  0.8 &  2.1 &  6.6 &  0.688 & 0.049   & 3.4 &  0.906 & 0.036   &  0.8 & $R$	\\
14 & 17:25:18.87 --26:46:07.79 & 40  &  4.2 &  1.7 &  5.1 &  0.669 & 0.047   & 2.5 &  1.237 & 0.049   &  0.5 & $R$	\\
15 & 17:36:57.59 --24:18:06.02 & --  &  0.3 &  1.4 &  4.9 &  0.469 & 0.034   & 0.5 &  1.346 & 0.054   &  0.3 & $R$	\\
16 & 17:14:56.44 --27:31:09.26 & --  &  0.5 &  1.5 &  4.1 &  0.323 & 0.023   & 0.2 &  0.416 & 0.017   &  0.8 & Src 2\tablenotemark{g}	\\
17 & 17:37:35.42 --23:30:52.19 & --  &  1.5 &  3.7 &  7.9 &  0.270 & 0.020   & 1.3 &  0.213 & 0.009   &  1.3 &  	\\
18 & 17:11:29.31 --27:25:36.30 &  9  & 19.4 & 12.4 & 15.6 &  0.246 & 0.018   & --  &  0.104 & 0.004   &  2.4 & [BHB2007] 15    \\
19 & 17:20:18.85 --26:55:57.35 & 29  &  4.3 &  1.4 &  4.8 &  0.241 & 0.017   & 1.8 &  0.014 & 0.001   & 16.7 & $R$	\\
20 & 17:34:01.76 --25:24:10.73 & 92  &  1.0 &  1.7 &  9.3 &  0.213 & 0.015   & 1.4 &  0.013 & 0.001   & 16.2 & $R$	\\
21 & 17:11:14.45 --27:26:54.40 &  9  & 19.4 &  7.5 & 11.8 &  0.210 & 0.016   & --  &  0.155 & 0.006   &  1.4 & [BHB2007] 4     \\
22 & 17:23:35.15 --23:42:24.05 & 100 &  2.4 &  2.8 &  4.1 &  0.184 & 0.013   & 0.8 &  0.011 & 0.001   & 16.8 & $R$	\\
23 & 17:11:11.78 --27:26:55.17 & 9   & 19.4 &  9.0 & 12.9 &  0.110 & 0.009   & 1.4 &  0.095 & 0.004   &  1.2 & [BHB2007] 3     \\
24 & 17:20:54.17 --26:46:42.77 & 34  &  9.3 &  2.3 &  5.7 &  0.110 & 0.009   & 0.7 &  0.014 & 0.001   &  7.6 &  	\\
25 & 17:29:28.24 --25:56:25.08 & 57  &  2.8 &  3.9 &  8.6 &  0.104 & 0.008   & 2.6 &  0.014 & 0.001   &  7.6 & $R$	\\
26 & 17:19:41.25 --26:55:31.67 & 29  &  4.3 &  8.1 & 10.6 &  --    & --      & --  &  0.163 & 0.007   &   -- &  	\\
\enddata
\tablecomments{Sources sorted according to their 70~$\mu$m flux. Photometry for Barnard~59 sources is from \citet{bro07}.}
\tablenotetext{a}{from MIPS 24\,$\mu$m data except for sources 8, 18 and 21, which were added from \citet{bro07}, see text}
\tablenotetext{b}{optical extinction from an extinction map with extended structure removed \citep{alv07}}
\tablenotetext{c}{optical extinction from an extinction map including extended structure \citep{lom06}}
\tablenotetext{d}{distance between MIPS70 and MIPS24 positions in source detection, possibly misleading for bright sources}
\tablenotetext{e}{$R$ = rejected candidate YSO (see text)}
\tablenotetext{f}{at MIPS wavelengths: blend with neighboring [BHB2007] 2; pair also known as LkH$\alpha$~346 AB.}
\tablenotetext{g}{in Covey et al. 2009 (in prep.)}
\end{deluxetable}

\begin{deluxetable}{rrrrrrrrrr}
\tabletypesize{\scriptsize}
\tablecaption{Candidate young stellar objects in the Pipe Nebula: 2MASS counterparts\label{tab_candsel2}}
\tablecolumns{8}
\tablehead{
\colhead{no.} & \colhead{2MASS name} & \colhead{2MASS-\textsl{J}} & \colhead{err.} & 
\colhead{2MASS-\textsl{H}} & \colhead{err.} & \colhead{2MASS-\textsl{K$_S$}} & \colhead{err.} \\
\colhead{} & \colhead{} & \colhead{[mag]} & \colhead{[mag]} & \colhead{[mag]} & \colhead{[mag]} & \colhead{[mag]} & 
\colhead{[mag]}
}
\startdata
1  & J17112317-2724315 & $>$18.78&   -- & $>$17.80&   -- & 15.08 & 0.14 \\
2  & J17111726-2725081 &   13.62 & 0.04 &   10.82 & 0.04 &  8.77 & 0.03 \\
3  & 		       &      -- &   -- &      -- &   -- &    -- &   -- \\
4  & J17110392-2722551 &   10.46 & 0.03 &    8.99 & 0.03 &  7.76 & 0.02 \\
5  & J17100811-2715190 &    9.07 & 0.03 &    7.23 & 0.04 &  5.80 & 0.03 \\
6  & J17112153-2727417 &   12.74 & 0.02 &   10.57 & 0.02 &  8.98 & 0.03 \\
7  & J17112701-2723485 &   11.88 & 0.04 &   10.14 & 0.04 &  9.08 & 0.03 \\
8  & J17111827-2725491 &$>$18.4  &   -- &   15.12 & 0.08 & 11.95 & 0.03 \\
9  & J17113036-2726292 &   11.91 & 0.03 &   10.00 & 0.03 &  8.89 & 0.02 \\
10 & J17350225-2523503 &   15.55 & 0.07 &   14.34 & 0.08 & 13.11 & 0.06 \\
11 & J17145108-2722399 &   10.26 & 0.02 &    8.89 & 0.02 &  8.16 & 0.02 \\
12 & J17112729-2725283 &   13.17 & 0.03 &   10.65 & 0.03 &  9.11 & 0.02 \\
13 & J17294185-2629092 &   13.75 & 0.04 &   10.06 & 0.02 &  7.63 & 0.02 \\
14 & J17251886-2646099 & $>$16.05&   -- &   13.23 & 0.04 &  9.72 & 0.03 \\
15 & J17365757-2418066 & $>$12.71&   -- &   10.65 & 0.06 &  8.40 & 0.03 \\
16 & J17145648-2731080 &   10.42 & 0.02 &    9.56 & 0.03 &  9.04 & 0.02 \\
17 & J17373543-2330517 &   14.60 & 0.05 &   12.38 & 0.04 & 10.92 & 0.03 \\
18 & J17112942-2725367 &   13.28 & 0.02 &   11.74 & 0.02 & 10.70 & 0.02 \\
19 & 		       &      -- &   -- &      -- &   -- &    -- &   -- \\
20 & J17340187-2524070 &   15.71 & 0.08 &   14.02 & 0.07 & 13.31 & 0.06 \\
21 & J17111445-2726543 &   11.62 & 0.02 &   10.42 & 0.02 &  9.63 & 0.02 \\
22 & J17233513-2342249 &   16.10 & 0.11 &   15.06 & 0.10 & 14.39 & 0.10 \\
23 & J17111182-2726547 &   14.09 & 0.03 &   12.62 & 0.03 & 11.76 & 0.03 \\
24 & J17205433-2646434 &   12.30 & 0.02 &   11.02 & 0.02 & 10.46 & 0.03 \\
25 & J17292824-2556252 &    9.01 & 0.03 &    7.93 & 0.03 &  7.47 & 0.03 \\
26 & J17194124-2655317 & $>$18.35&   -- & $>$15.86&   -- & 14.57 & 0.08 \\
\enddata
\tablecomments{Sources 3 and 19 do not have 2MASS counterparts.} 
\end{deluxetable}

\begin{deluxetable}{lrrr}
\tabletypesize{\scriptsize}
\tablecaption{Comparison of detections in the high-extinction regions and the control regions\label{tab_statcomp}}
\tablewidth{0pt}
\tablecolumns{4}
\tablehead{
\colhead{} & \colhead{control region} & \colhead{scaled to high-ext. area} & \colhead{high-ext. region}\\
\colhead{} & \colhead{N} & \colhead{N} & \colhead{N}\\
}
\startdata
"class I"    &      48        &    2.9	       &    9	     \\ 
"class II"   &      64        &    4.0	       &   16	     \\ 
"class III"  &      87        &    5.2	       &    9	     \\
\enddata
\tablecomments{The designations in the first column correspond to the flux space regions shown in Figures~\ref{fig_selyso} and~\ref{fig_selbkg}.}
\end{deluxetable}

\begin{deluxetable}{rrrrrrr}
\tabletypesize{\scriptsize}
\tablecaption{Fit results for candidate young stellar objects in the Pipe Nebula\label{tab_candfit}}
\tablewidth{0pt}
\tablecolumns{7}
\tablehead{
\colhead{Source} & \colhead{$A_V$(tot)} &               & \colhead{$L_{\rm bol}$} &                     & \colhead{$\alpha_{\rm K-24}$} & \colhead{class\tablenotemark{a}} \\
                 & \colhead{mag}        & \colhead{mag} & \colhead{$L_\odot$}     & \colhead{$L_\odot$} &                               &                 \\
                 & \colhead{min}        & \colhead{max} & \colhead{min}           & \colhead{max}       &                               &                 
}
\startdata
1   & 116.7 & 179.9 &	1.4 &	2.6  &   2.71 & I  \\	  
2   &  20.7 & 117.7 &  16.5 &  30.1  &   0.58 & I  \\	  
3   &  78.2 & 996.9 &	0.7 &  20.1  & --     & I  \\	  
4   &   8.1 & 106.3 &	7.0 &  29.6  &   0.03 & f \\	  
5   &   8.7 &  62.9 &  55.9 & 175.4  & --0.73 & II \\	  
6   &  14.1 &  24.7 &	2.4 &  11.5  & --0.09 & f \\	  
7   &   8.9 &  12.8 &	0.7 &	2.7  & --0.30 & f \\	  
8   &  40.1 &  48.1 &	1.4 &  12.3  &   0.50 & I  \\	  
9   &   9.3 &  13.8 &	0.5 &	4.5  & --1.09 & II \\	  
11  &   4.3 &  34.0 &	0.8 &	6.6  & --0.83 & II \\	  
12  &  15.3 &  22.4 &	1.0 &	7.6  & --0.54 & II \\	  
16  &   1.9 &  14.8 &	0.6 &	2.5  & --0.61 & II \\	  
17  &  15.0 &  16.4 &	0.3 &	0.6  & --0.17 & f \\	  
18  &   6.6 &  36.7 &	0.1 &	0.8  & --0.55 & II \\	  
21  &   4.5 &  21.9 &	0.2 &	1.1  & --0.79 & II \\	  
23  &   9.5 &  26.4 &	0.1 &	0.2  & --0.18 & f \\	  
24  &   1.1 &	3.2 &	0.1 &	0.2  & --1.47 & II \\	  
26  &  30.0 &  31.7 &	0.3 &	0.3  &   1.12 & I  \\	  
\enddata
\tablecomments{For source 9, the goodness-of-fit criterion has been relaxed to allow for more than one fit (see text).}
\tablenotetext{a}{evolutionary class, determined from the full SEDs; $\alpha_{K-24}$ is used to differentiate flat-spectrum from class I and II sources \citep{gre94}: $\alpha \ge 0.3\to$\, class I; $-0.3\le \alpha<0.3\to$\, class f (flat spectrum); and $\alpha<-0.3\to$\, class II}
\end{deluxetable}

\clearpage

\begin{figure*}
\centering
\includegraphics[width=0.9\linewidth]{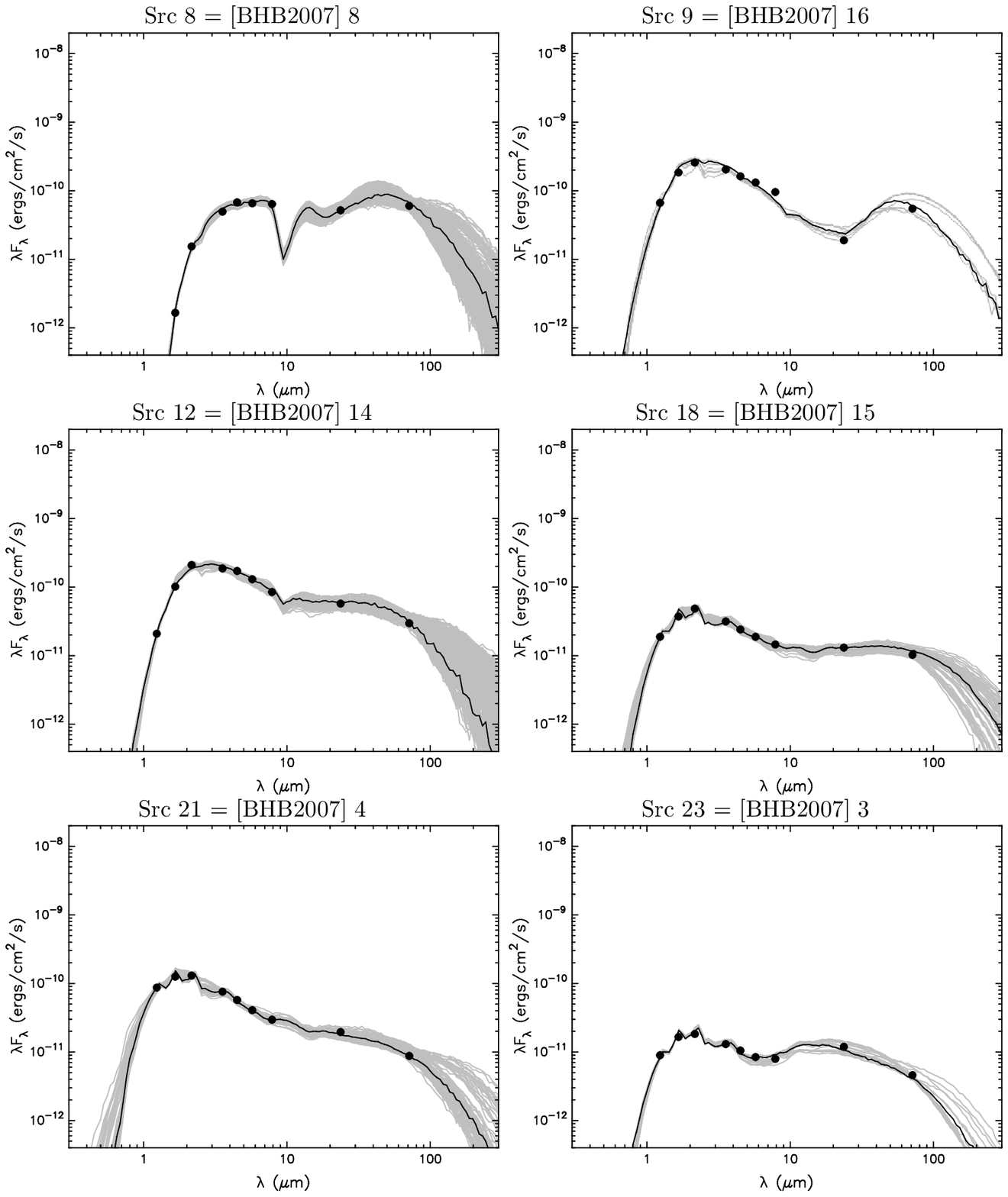}

{Fig.~\ref{fig_seds1} -- cntd.}
\end{figure*}

\begin{figure*}
\centering
\includegraphics[width=0.9\linewidth]{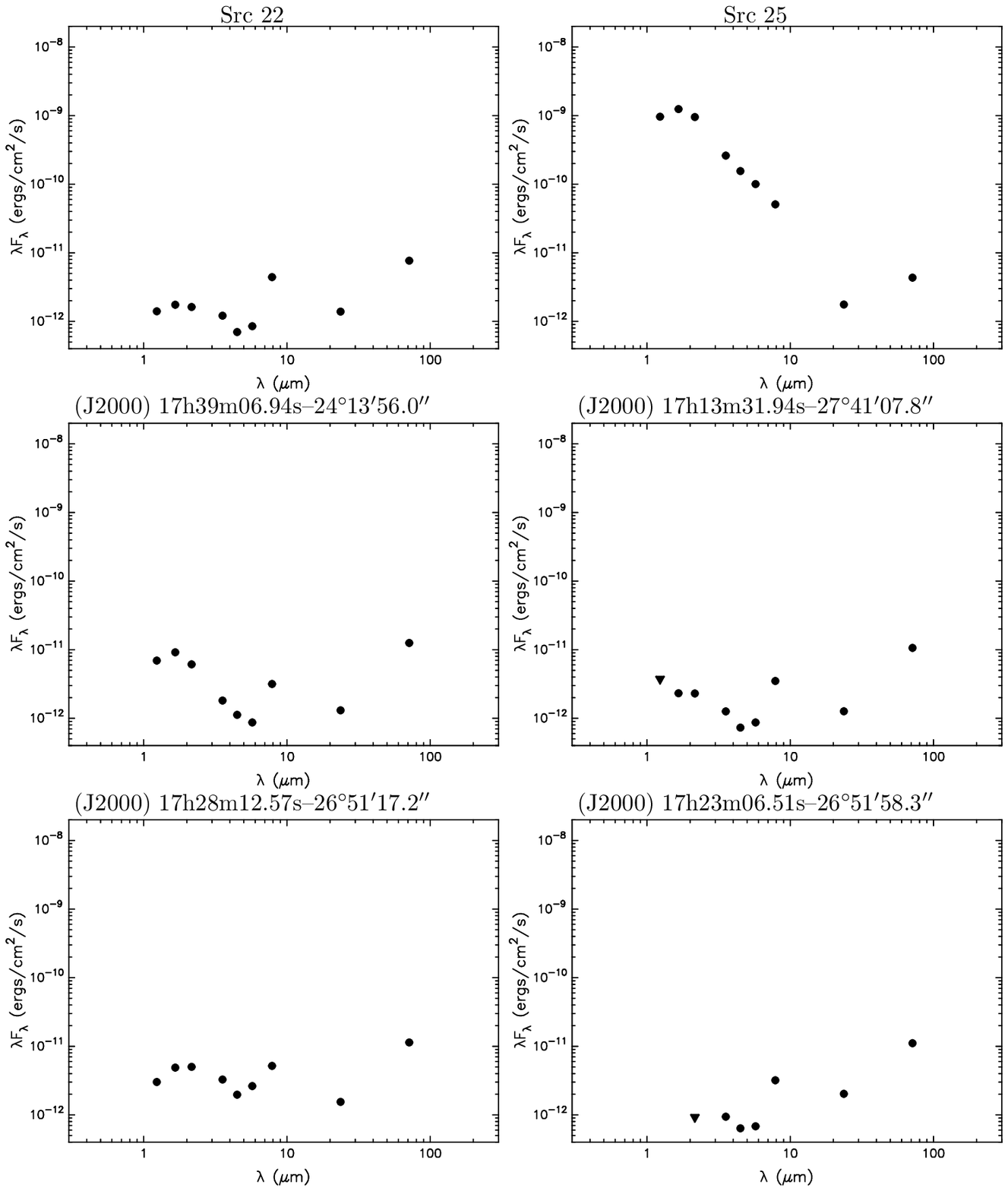}

{Fig.~\ref{fig_seds3} -- cntd.}
\end{figure*}

\end{document}